\documentclass[a4paper,12pt]{article}

\usepackage[english]{babel}
\usepackage[utf8x]{inputenc}
\usepackage[T1]{fontenc}
\usepackage{url}
\usepackage{authblk}
\usepackage{newtxtext,newtxmath} 

\usepackage{geometry}
\usepackage{setspace}
\usepackage[numbers,sort&compress]{natbib}

\usepackage{graphicx}
\usepackage{caption}
\DeclareCaptionLabelSeparator{bar}{\,$\mid$\,}
\usepackage{amsmath} 
\usepackage{amssymb} 
\usepackage[detect-all]{siunitx}

\begin{document}
\renewcommand{\abstractname}{\vspace{-\baselineskip}}

\title{Full phase diagram of isolated skyrmions in a ferromagnet} 
\author[]{Felix B\"{u}ttner\thanks{felixbuettner@gmail.com} }
\author[]{Ivan Lemesh}
\author[]{Geoffrey S. D. Beach}
\affil[]{Department of Materials Science and Engineering, Massachusetts Institute of Technology, Cambridge, Massachusetts 02139, USA}

\date{\today}

\maketitle

\begin{abstract}
\normalsize
Magnetic skyrmions are topological quasi particles of great interest for data storage applications because of their small size, high stability, and ease of manipulation via electric current. Theoretically, however, skyrmions are poorly understood since existing theories are not applicable to small skyrmion sizes and finite material thicknesses. Here, we present a complete theoretical framework to determine the energy of any skyrmion in any material, assuming only a circular symmetric \SI{360}{\degree} domain wall profile and a homogeneous magnetization profile in the out-of-plane direction. Our model precisely agrees with existing experimental data and micromagnetic simulations. Surprisingly, we can prove that there is no topological protection of skyrmions.  We discover and confirm new phases, such as bi-stability, a phenomenon unknown in magnetism so far. The outstanding computational performance and precision of our model allow us to obtain the complete phase diagram of static skyrmions and to tackle the inverse problem of finding materials corresponding to given skyrmion properties, a milestone of skyrmion engineering.
\end{abstract}

\maketitle 

\section*{Introduction}

Magnetic skyrmions are spin configurations with spherical topology \cite{nagaosa_topological_2013,buttner_magnetic_2016}, typically manifesting as circular domains with defect-free domain walls (DWs) in systems with otherwise uniform out-of-plane magnetization. Skyrmions are the smallest non-trivial structures in magnetism and they behave like particles \cite{tomasello_strategy_2014,makhfudz_inertia_2012,everschor_current-induced_2011,iwasaki_current-induced_2013,sampaio_nucleation_2013,buttner_dynamics_2015}, which makes them of fundamental interest and of practical utility for high-density data storage applications \cite{fert_skyrmions_2013,wiesendanger_nanoscale_2016,rosch_skyrmions:_2013}. Skyrmions have been investigated for many decades \cite{malozemoff_magnetic_1979}, but only recently has attention shifted to the detailed DW structure and thereby-determined topology. Two factors have driven this trend: technological advances allowing for direct imaging of the spin structure \cite{heinze_spontaneous_2011,yu_real-space_2010,romming_field-dependent_2015} and the discovery that the Dzyaloshinskii-Moriya interaction (DMI) can be used to stabilize that structure. In particular, DMI can lead to a skyrmion global ground state above the Curie temperature ($T_c$) in a Ginzburg-Landau theory of a ferromagnet \cite{rosler_spontaneous_2006}. While stray fields and external fields are not included in that model, they can further stabilize skyrmions up to room temperature (below $T_c$), as observed in thin film multilayers \cite{moreau-luchaire_additive_2016,boulle_room-temperature_2016,woo_observation_2016,buttner_dynamics_2015,yu_room-temperature_2016,jiang_blowing_2015}. 

While homochiral skyrmions were first observed in bulk materials with broken inversion symmetry, multilayers stacks, such as such as Pt/Co/Ir \cite{moreau-luchaire_additive_2016}, Ta/CoFeB/TaO$_x$ \cite{jiang_blowing_2015,jiang_direct_2016}, Pt/Co/Ta \cite{woo_observation_2016}, and Pt/CoFeB/MgO \cite{woo_observation_2016,litzius_skyrmion_2016} with arbitrary repetitions of these layers, have seen increasing popularity recently.  In such structures, inversion symmetry breaking and spin-orbit coupling at the ferromagnet/heavy-metal interface can lead to a strong DMI that promotes magnetic skyrmions with well-defined chirality. Multilayers are particularly attractive since the relevant energy terms (e.g., anisotropy, DMI, magnetostatic) can be engineered by controlling the interfaces and the volume fraction of magnetic material in the stack.  However, with this flexibility comes a considerable engineering challenge in that the parameter space for engineering is overwhelmingly large: It has six dimensions (five material parameters plus external magnetic field) and thanks to the interfacial origin of many magnetic properties and to the effective medium scaling of these properties with the thickness of non-magnetic spacer layers \cite{woo_observation_2016,lemesh_accurate_????}, all of these dimensions can be tuned individually. Blind experimental investigation of this parameter space is hence prohibitive. Existing theories are of limited help, since they either involve crude approximations that fail to reproduce the wealth of skyrmion states even qualitatively \cite{kooy_experimental_1960,malozemoff_magnetic_1979,ezawa_giant_2010,guslienko_skyrmion_2015,rohart_skyrmion_2013} or contain unsolved complicated integrals and differential equations, which renders the evaluation of the theory extremely computationally expensive and slow (for instance, micromagnetic simulations and Refs.~\cite{tu_determination_1971,bogdanov_thermodynamically_1994,kiselev_chiral_2011}). In particular, most theories ignore the non-local nature of stray field interactions \cite{bogdanov_thermodynamically_1994,kiselev_chiral_2011,leonov_properties_2016,rohart_skyrmion_2013}, which is responsible for many interesting features in the intermediate film thickness regime. A single coherent theory  that quickly and accurately predicts the existence and the properties of isolated skyrmions for any given point in the six-dimensional parameter space remains elusive. Here, we provide a theoretical model  for all energy terms of an isolated skyrmion in a given material with infinite in-plane extent. The model is fully analytical and accurate to \SI{1}{\percent} in the entire parameter space. Thanks to the analytical nature, we can find energy minima extremely quickly by searching for roots of the partial derivatives. The resulting equilibrium states show excellent agreement with simulations and experiments. Using our theory, we find exotic new states, such as multi-stabilities (co-existence of skyrmions with different properties in the same sample under the same conditions), zero stiffness skyrmions, and zero field skyrmions. These new states can have many novel applications, some of which we suggest here. We obtained the minimum energy skyrmion states for millions of material parameter and field combinations in less than a week, yielding the full phase diagram of skyrmion states and demonstrating that our model is suitable to solve the inverse problem of engineering skyrmion properties through material selection.

Our theory takes as input the uniaxial anisotropy constant $K_u$, saturation magnetization $M_s$, exchange constant $A$, interface and bulk DMI strengths $D_i$ and $D_b$, magnetic layer thickness $d$, and applied out-of-plane field $H_z$. Given these parameters, we derive the energy function that determines the spin structure of skyrmions in any material. In general, stable skyrmions correspond to sufficiently deep minima of the energy functional $E[\mathbf{m}]$ of all possible spin structures $\mathbf{m}(\mathbf{r})$, where $\mathbf{m}$ is the unit magnetization vector and $\mathbf{r}$ is the position vector.  In practice, minimizing the energy functional in its raw form, as done in micromagnetic simulations, is prohibitively slow for systematic skyrmion engineering. Our simple and efficient analytical model is enabled by the recent experimental confirmation \cite{romming_field-dependent_2015,boulle_room-temperature_2016} of an analytic and universal \SI{360}{\degree} DW model \cite{braun_fluctuations_1994} for the spin structure $\mathbf{m}(\mathbf{r})$ of all skyrmions. The full analytic model for the total energy function is provided in the supplemental information, along with a detailed discussion of how to solve the integrals of the individual interactions. Here, we focus on its implications.

\section*{Effective energy contributions}

\begin{figure}
\centering
\includegraphics[width=\textwidth]{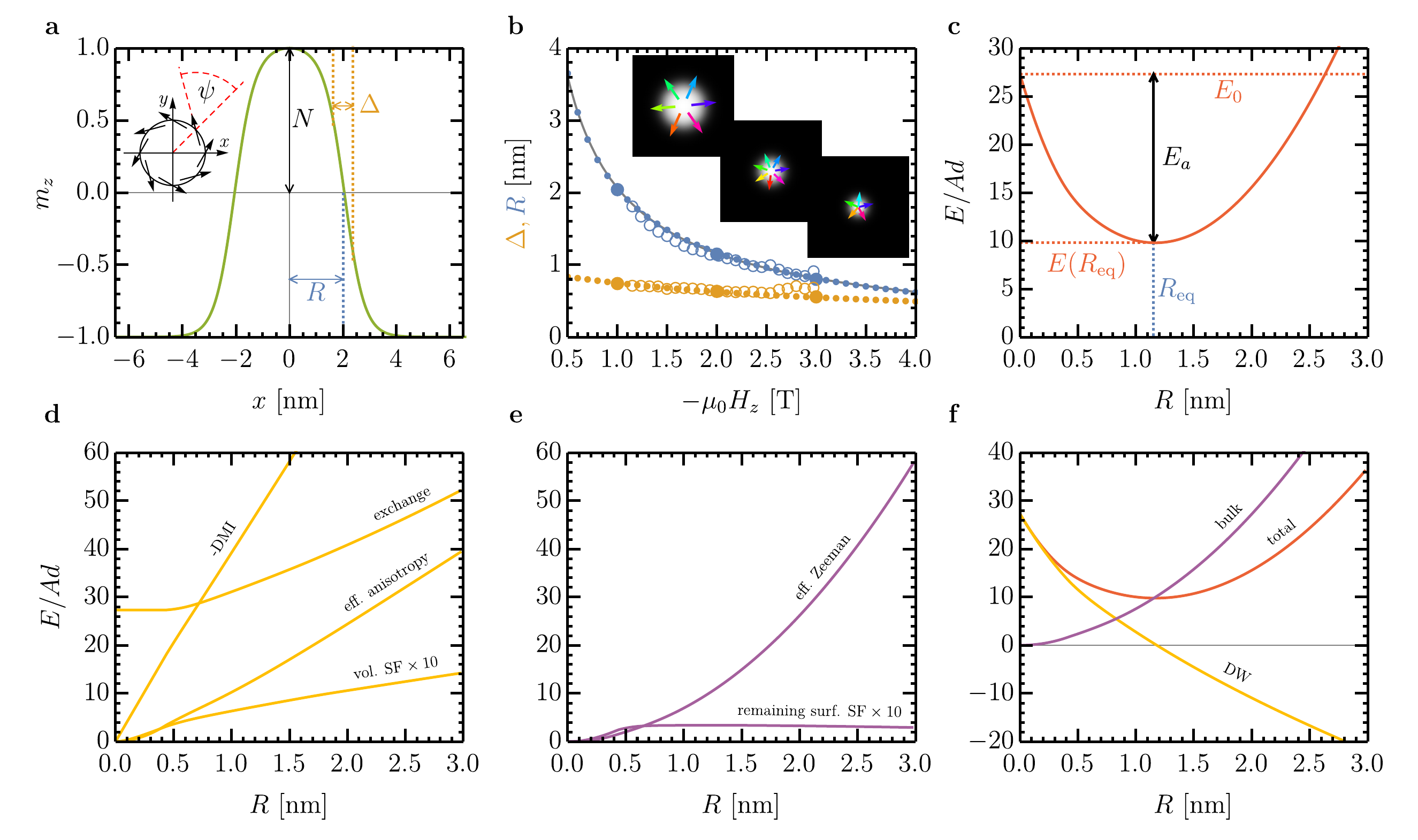}
\caption{\textbf{Skyrmion profile and individual energy terms.} \textbf{a}, Normalized perpendicular magnetization $m_z$ as a function of position $x$ along the diameter of a skyrmion and definition of the characteristic parameters $R$ (radius), $\Delta$ (DW width), $\psi$ (DW angle, inset) and $N$. The profile corresponds to the $\mu_0H_z=\SI{-1}{T}$ data point in \textbf{b}. \textbf{b}, Radius and DW width as a function of applied magnetic field $\mu_0H_z$. The small solid data points are predictions of our analytical model, the solid large data points are results from micromagnetic simulations, the open points are experimental results of Romming \textit{et al.} \cite{romming_field-dependent_2015}, and the solid grey line is a fit with the simple model of Eq.~\eqref{eq:RvsHzFit}. The insets show the relaxed spin structures obtained from micromagnetic simulations corresponding to the large solid data points. \textbf{c}, Total energy $E$, normalized to $Ad$, as a function of skyrmion radius at a field of $\mu_0H_z=\SI{-2}{T}$. At each point of the solid red line the energy has been minimized with respect to $\Delta$ and $\psi$ while keeping $R$ fixed. $R_\text{eq}$ is the equilibrium radius as plotted in \textbf{b}. \textbf{d-f}, Decomposition of the total energy in \textbf{c} into individual components. \textbf{d}, DW energies: DMI energy (inverted), exchange energy, effective anisotropy energy, and volume stray field energy (multiplied by 10). \textbf{e}, Bulk energies: effective Zeeman energy and remaining surface stray field energies (multiplied by 10). \textbf{f}, Sum of all DW and sum of all bulk energies, together with the total energy. In this particular example, the sum of DW energies has a negative slope at the minimum of the total energy, which makes this skyrmion DMI stabilized according to our definition. The parameters for this data set are experimental values from Ref.~\cite{romming_field-dependent_2015} and our predictions are in excellent agreement with the experimental observations.}
\label{fig:1}
\end{figure}

The skyrmion spin structure (Fig.~\ref{fig:1}a) can be accurately described by four parameters: radius $R$, DW width $\Delta$, DW angle $\psi$, and topological charge $N$. $R$ and $\Delta$ are independent parameters that determine the magnetization profile $m_z(x,y)$, whereas $\psi$ specifies whether the in-plane component of the DW spins is radial (N\'eel, $\psi=0,\pi$), azimuthal (Bloch, $\psi=\pi/2,3\pi/2$), or intermediate (transient). For large $\rho=R/\Delta$, skyrmions consist of an extended out-of-plane magnetized domain bounded by a narrow circular DW, while for $\rho \sim 1$ the inner domain is reduced to a point-like core resembling a magnetic vortex.  We refer to these limiting cases as bubble skyrmions and vortex skyrmions, respectively, consistent with the literature, but note that many skyrmions observed recently \cite{moreau-luchaire_additive_2016,yu_variation_2016,woo_observation_2016,buttner_dynamics_2015} showed intermediate values of $\rho$ and cannot be classified distinctly.

The case of bubble skyrmions is readily treated analytically through the so-called wall-energy model known for a long time \cite{cape_magnetic_1971}.  In this limit, the skyrmion energy simplifies to
\begin{align}
E=2\pi d \sigma_\text{DW} R+aR-bR\ln(R/d)+cR^2\label{eq:large_radius_energy}
\end{align}
with $2\pi d \sigma_\text{DW} R$ being the DW energy ($\sigma_\text{DW}$ is the energy density of an isolated DW), $aR+bR\ln(R/d)$ the Zeeman-like surface stray field energy, and $cR^2$ the Zeeman energy.  Here $a,b,c$ are material-dependent parameters that include corrections to the original model \cite{cape_magnetic_1971} to account for DMI, volume charges, and finite $\Delta$ (see supplemental information). The crucial assumption of Eq.~\eqref{eq:large_radius_energy} is that $\sigma_\text{DW}$, $\Delta$, and $\psi$ do not depend on $R$ and that hence $\Delta$ and $\psi$ can be obtained from minimizing $\sigma_\text{DW}$ while $R$ is the minimum of $E$ assuming a constant $\sigma_\text{DW}$. The simplicity of this wall-energy model is reason for its popularity \cite{rohart_skyrmion_2013,schott_skyrmion_2016}. However, our accurate model shows that $\Delta$ can change with $R$ even up to unexpectedly large radii of $R=100\Delta$. Indeed, we find that the wall-energy model fails quantitatively and qualitatively for almost all skyrmions that are of interest today. Our model fully accounts for the correlation between radius, DW width and DW angle, thereby providing a single theoretical framework that accurately describes any skyrmion. Importantly, our model reveals unexpected and qualitatively new exotic behaviors that are precluded by the approximations inherent in prior treatments.

Our analytic equations for the total energy function can be numerically minimized with respect to $R$, $\Delta$, and $\psi$, for a given set of material parameters and external field $H_z$, to obtain the equilibrium skyrmion configuration. The predictions of our model agree precisely with micromagnetic simulations and with the experimental data of Romming \textit{et al.} \cite{romming_field-dependent_2015}, see Fig.~\ref{fig:1}b. Note that fields are negative in our convention, antiparallel to the skyrmion core. Heuristically, we find that the function
\begin{align}
R(H_z)\approx \frac{a_1}{(-H_z)^{a_2}}+a_3\label{eq:RvsHzFit}
\end{align}
fits the field dependence of the equilibrium radius for a wide range of parameters.

Our model yields the energy of a given skyrmion configuration in less than a millisecond on a regular personal computer, thereby providing dramatic improvement over micromagnetic simulations in terms of computation speed. Moreover, it gives access to information that cannot be obtained by simulations.  For example, by virtue of taking partial derivatives, the model allows to minimize $\Delta$ and $\psi$ for any non-equilibrium skyrmion radius and, therefore, to obtain the energy as a function of radius $E(R)$. Micromagnetic simulations can only yield the equilibrium $R,\Delta,\psi$. The $E(R)$ curve directly relates to the skyrmion stability (by quantifying the energy barriers) and to its rigidity (by the curvature near the energy minimum). Figure~\ref{fig:1}c shows $E(R)$ calculated for the skyrmion described in Fig.~\ref{fig:1}b at $\mu_0 H_z=$-2T, which exhibits a single minimum corresponding to an isolated skyrmion. The only other stable state is the ferromagnetic ground state at $E=0$. Despite the different topology of the skyrmion and the ferromagnetic state, there is a continuous path from one to the other, which goes through the singular $R=0$ state. The singluar $R=0$ state does not have a topology, which is why topological quantization is lifted here. Remarkably, the energy along the path towards $R=0$ remains finite, in contrast to earlier believes \cite{rohart_skyrmion_2013}. At $R=0$, the skyrmion energy takes a universal value,
\begin{align}
E(R=0)=E_0=27.3Ad.\label{eq:27Ad}
\end{align}
The zero radius energy depends only on $A$ and $d$ and not on DMI. By finding a topologically valid and energetically possible path to annihilation we prove that skyrmions are not protected by their topology, even in continuum models and in the presence of strong DMI. Note that a finite value for the energy barrier has been found before \cite{leonov_properties_2016,belavin_metastable_1975}, but the role of DMI and implications for the topological stability were not discussed. The skyrmion of Fig.~\ref{fig:1}c exhibits a finite annihilation energy $E_a = E_0-E(R)$ and a nucleation energy barrier $E_n = E_0$. Note that $E_n$ and $E_a$ overestimate the energy barriers because skyrmions can deform in a way that is not covered by the \SI{360}{\degree} DW model underlying our calculations and therefore reduce the nominal energy barrier \cite{rohart_path_2016}. However, previous studies \cite{rohart_path_2016} and our own micromagnetic simulations indicate that the reduction of the energy barrier due to deformations is smaller than $2Ad$ (even though sometimes extremely small cell sizes are required). This is in excellent agreement with the observation of Belavin and Polyakov \cite{belavin_metastable_1975} that the minimum energy of a skyrmion state in a model that includes only exchange energy is $8\pi Ad$, which is approximately $2Ad$ smaller than the zero radius exchange energy of our \SI{360}{\degree} DW theory. Including thermal effects, we therefore consider minima of the total energy to be stable in our discussions below if $E_a>2Ad+k_BT$.

All energy contributions to the total energy can be classified into two categories, DW and bulk energies, and inspection of the individual energy terms allows one to identify the mechanism responsible for skyrmion stability. At large radii, DW energies are linear in $R$, whereas all non-linear terms are bulk energies. Exchange, anisotropy, DMI, and volume stray field energies are DW energies. The Zeeman energy is a bulk energy. Surface stray fields contribute to both categories: The DW contribution leads to an effective reduction of the anisotropy and the bulk contribution effectively reduces the external field.

The decomposition of the total energy into DW and bulk terms is depicted in Figs.~\ref{fig:1}d-f. Minima in $E(R)$ can exist only if energy terms with positive slope are compensated by terms with negative slope, and the latter can only arise through DMI and surface stray fields.  One can therefore classify skyrmions as DMI (stray field) stabilized if the sum of DW (bulk) energies has a negative slope at the equilibrium radius $R_\text{eq}$.  Our model hence provides the first mathematical basis for a terminology commonly used in the literature without rigorous justification \cite{nagaosa_topological_2013}. 

\section*{New phases}

\begin{figure}
\centering
\includegraphics[width=\textwidth]{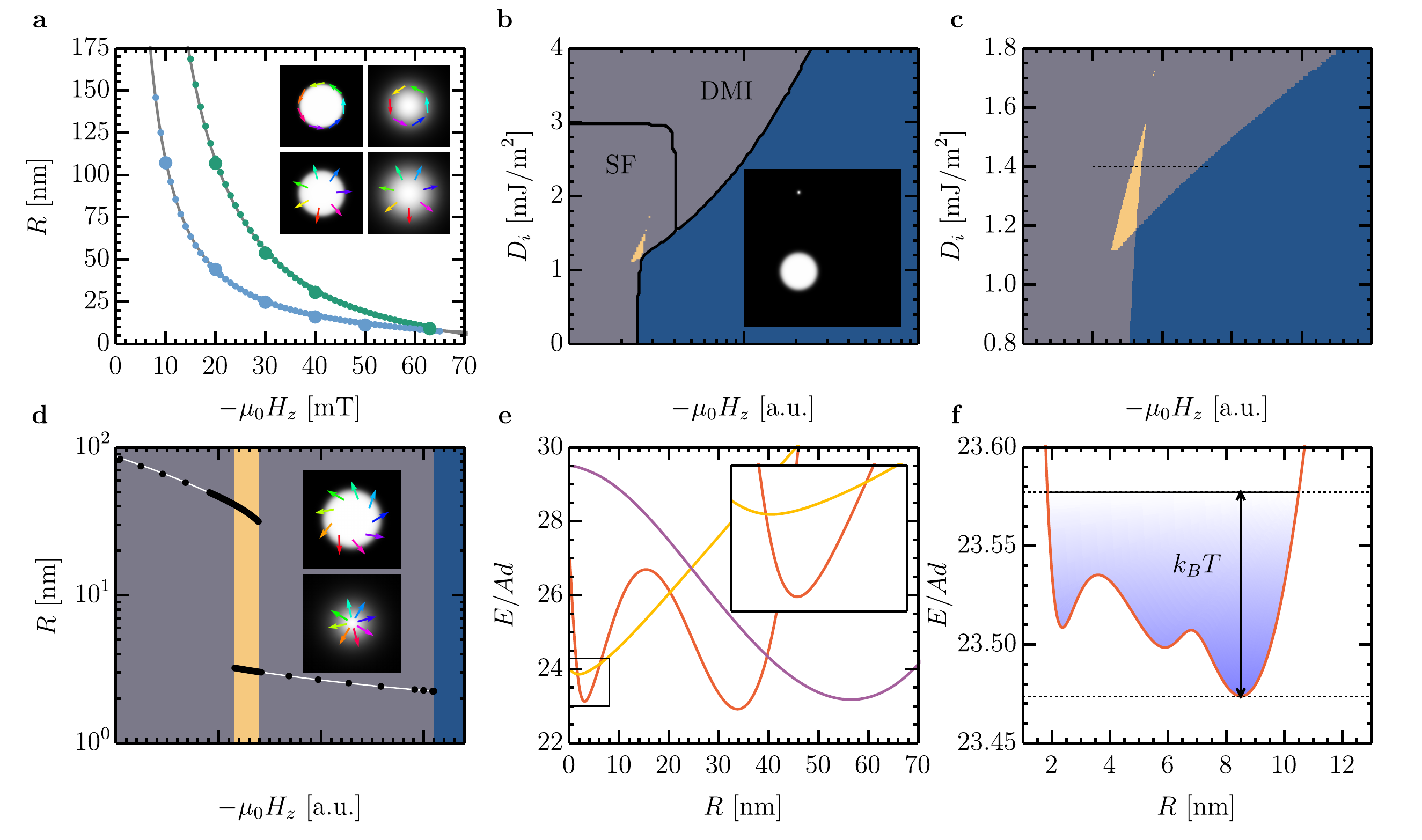}
\caption{\textbf{Richness of the skyrmion phase diagram.} \textbf{a}, A Bloch (green) and a N\'eel (blue) skyrmion with very similar radii as a function of applied field. Small data points are predictions of our model, large data points are extracted from micromagnetic simulations, and the grey solid lines are fit to Eq.~\eqref{eq:RvsHzFit}. The inset images show the skyrmions at the smallest and the largest simulated field magnitude on the left and right, respectively. Images at the largest field are magnified by a factor 10. The top inset row shows the spin structures of the non-DMI skyrmion and the bottom row refers to the DMI system. \textbf{b}, Phase diagram of the multi-stability of skyrmions as a function of applied field and DMI strength, where blue, grey, and yellow indicate the regions of instability, mono-stability, and bi-stability, respectively. The solid black line is the border between stray-field stabilized skyrmions (SF) and DMI stabilized skyrmions (DMI). The inset shows a simulation of two simultaneously present distinct skyrmions in the bi-stability regime. \textbf{c}, Magnification of the bi-stability region. The dashed horizontal line indicates the slice that is plotted in panel \textbf{d}. \textbf{d}, Radius as a function of applied field for skyrmions in the bi-stability region in \textbf{c} at $D_i=\SI{1.4}{mJ/m^2}$. The solid white lines are fit to Eq.~\eqref{eq:RvsHzFit}. The background color indicates the level of multi-stability as plotted in \textbf{c}. The insets show simulations of two types of skyrmions (large skyrmion on top, small skrymion at the bottom, not to scale). The small skyrmion is of pure N\'eel type whereas the large skyrmion represents an intermediate state between Bloch and N\'eel DW spin orientation. \textbf{e}, Total energy (red), DW energies (orange), and bulk energies (purple) as a function of radius. DW and bulk energies were scaled by $1/20$ and off-set artificially to fit into the plot area. The inset shows a magnification of the region marked by the black rectangle in the main panel. \textbf{f}, Energy as a function of radius for a zero stiffness skyrmion. The thermal energy available to the skyrmion at room temperature is indicated by the blue shading and the dotted lines. All states between $R=\SI{2}{nm}$ and $R=\SI{11}{nm}$ are accessible by thermal excitation.}
\label{fig:2}
\end{figure}

We now apply our model to gain further insight into skyrmion properties and to analyze the phase diagram of static isolated skyrmions. Features of particular interest are highlighted in Fig.~\ref{fig:2}. First, Fig.~\ref{fig:2}a compares the radius of a Bloch skyrmion in a material without DMI to a N\'eel skyrmion in a high DMI material (with $D_i>D_{c\psi}^\text{SW}$, where $D_{c\psi}^\text{SW}$ is the critical DMI value required to stabilize N\'eel DWs; see supplementary information). It is widely believed that small skyrmions exist only in materials with large DMI \cite{moreau-luchaire_additive_2016,kiselev_comment_2011,kiselev_chiral_2011,nagaosa_topological_2013,wiesendanger_nanoscale_2016}. Indeed, our results indicate that skyrmions with the sharpest core (quantitatively, values of $\rho< 1.3$) and skyrmions with sub-nanometer radius can only be found in materials with sufficiently large DMI. However, both skyrmions in Fig.~\ref{fig:2}a have a similar $R(H_z)$ dependence and collapse at a similar radius of less than \SI{10}{nm}. Near collapse, both skyrmions approach vortex-like configurations, which is agreement with recent experimental observations of vortex-like skyrmions in non-DMI materials \cite{yu_variation_2016}. The distinction between the two types is hardly possible based on their size with the resolution of most of the imaging tools available today and it is not possible to deduce the value of $\psi$ from measurements of $R$. 

Figures~\ref{fig:2}b,c depict the phase diagram of DMI-stabilized and stray-field-stabilized skyrmions. Surprisingly, $E(R)$ can exhibit multiple minima separated by energy barriers sufficiently high that they are individually stable, leading to a bi-stability (discussed in detail in the supplemental information). Bi-stability exists in a small pocket of the phase diagram, wherein two types of skyrmions can coexist in a material under identical conditions (Figs.~\ref{fig:2}b, c). We find this pocket in the stray-field stabilized part of the phase diagram where the phase boundary of the instability region has a cusp. The two types of skyrmions in the bi-stability region have very different properties (Fig~\ref{fig:2}d), confirmed by micromagnetic simulations: Their radii differ by more than one order of magnitude and their spin structure is N\'eel-like for the small skyrmion and transient for the large skyrmion. The transient value of $\psi$ for the larger skyrmion originates in a wider domain wall, which increases the importance of the volume stray field energy that favors a Bloch-like spin orientation. The different size and domain wall angle can be used to move the skyrmions in non-collinear directions by spin orbit torques, as detailed in the supplemental information.

The unexpected emergence of multiple minima in $E(R)$ is a consequence of introducing $\Delta$ and $\psi$ as free parameters, resulting in $\Delta(R)$ being nonlinear and sometimes nonmonotonic. The existence of degenerated isolated domain states is new in the entire field of magnetism. Decomposition of $E(R)$ into DW and bulk energies (Fig.~\ref{fig:2}e) reveals that the origin of this phenomenon is that each term individually exhibits a minimum. The minimum in $E(R)$ at small radius derives from the minimum in the DW energy terms, shifted towards larger radii by the negative-sloped bulk energies and vice versa for the large radius minimum. This observation helps explain why the phase boundaries between stray field stabilized and DMI stabilized skyrmions in Fig.~\ref{fig:2}b are vertical and horizontal (see also supplemental figure S1). The horizontal line marks the critical DMI value above which $\sigma_\text{DW}$ is negative, i.e., where the DW energies have a negative slope everywhere and all minima are DMI stabilized. The vertical line indicates the critical field value above which the applied field fully compensates the Zeeman-like surface stray field, meaning that the bulk energies are always positive with a positive slope beyond that point and again all minima are DMI stabilized. In any of these cases, either the DW or the bulk energies cease to have a minimum, which finally also explains why we find the bi-stable phase pocket in the stray field stabilized phase.

The last peculiar phenomenon we uncover in our analysis is the existence of zero stiffness skyrmions. Figure~\ref{fig:2}f shows $E(R)$ for a system that manifests three energy minima, where the maxima between these minima can easily be overcome by thermal energies at room temperature. In this particular example, the skyrmion radius can thermally fluctuate between \SI{2}{nm} and \SI{11}{nm}, such that it exhibits effectively zero stiffness with respect to variations of radius within this range. We expect that such skyrmions have a very low resonance frequency associated with their breathing mode, which could be exploited in skyrmion resonators \cite{schwarze_universal_2015} and should have impact on their inertia \cite{buttner_dynamics_2015} and on skyrmion Hall angle \cite{litzius_skyrmion_2016,jiang_direct_2016}.

\section*{Applications}

\begin{figure}
\centering
\includegraphics[width=\textwidth]{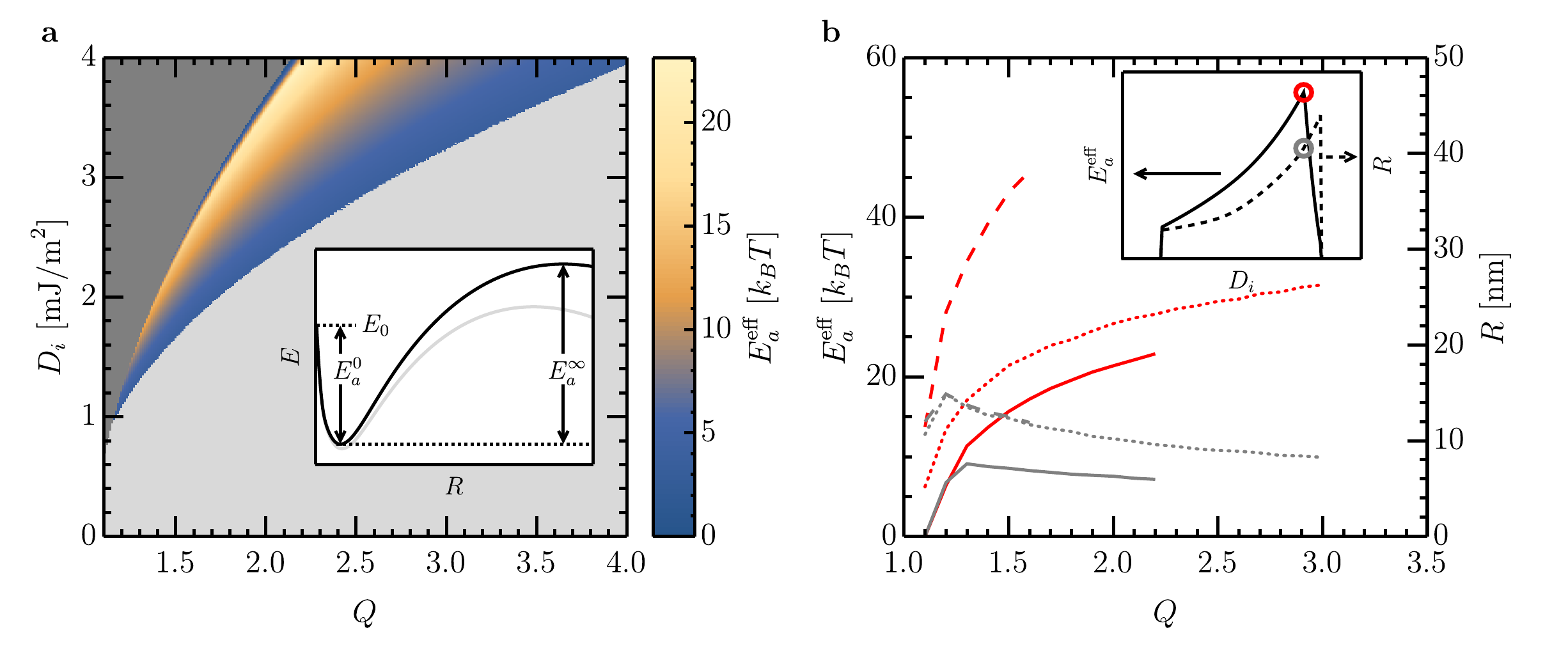}
\caption{\textbf{Zero field skyrmions.} \textbf{a}, Phase pocket for the existence of zero field skyrmions. The inset explains the two energy barriers, one towards smaller radii ($E_a^0$) and one towards larger radii ($E_a^\infty$), the overcoming of which leads to the collapse of the skyrmion. The grey line has a \SI{1}{\percent} larger DMI value compared to the black line. The total effective annihilation energy barrier $E_a^\text{eff}$, as color-coded in the main panel in units of the thermal energy $k_BT$, is the minimum of $E_a^0$ and $E_a^\infty$ minus the estimated internal energy of $2Ad$ that can be drawn from deformations, at given material parameters. The light grey area indicates that no minima exist at zero fields or that $E_a^\text{eff}$ is smaller than $2Ad+k_BT$. The dark grey area illustrates where spontaneous domain nucleation is expected at zero field, which is excluded from being useful for isolated skyrmion applications. \textbf{b}, The inset shows a slice of panel \textbf{a} at $Q=1.8$, plotting the annihilation energy barrier and the equilibrium skyrmion radius. The maximum energy barrier and the corresponding skyrmion radius are highlighted by a red and grey circle, respectively. This maximum energy barrier and corresponding radius are plotted in the main panel as a function of Q (effectively showing a slice of panel \textbf{a} when moving along the maxima of $E_a^\text{eff}$). The solid lines correspond to the regular material parameters used throughout this paper, the dotted lines correspond to a reduced $M_s$ and the dashed lines represent a larger exchange constant. All lines end when the maximum $E_a$ is found at $D_i>\SI{4}{mJ/m^2}$, which to the best of our knowledge is the maximum DMI that can be engineered. If larger DMI values were allowed, all energy lines would continue to go up and all radii lines would continue to go down.}
\label{fig:4}
\end{figure}

We now consider the design of skyrmions suitable for applications, such as racetrack-type memory devices in which bit sequences are encoded by the presence and absence of skyrmions that can be shifted by electric current \cite{parkin_magnetic_2008,fert_skyrmions_2013,wiesendanger_nanoscale_2016,rosch_skyrmions:_2013}. Three key attributes for such applications are (i) small bit sizes, (ii) long term thermal stability, and (iii) skyrmion stability in zero applied field. Indeed, we find a section in the phase diagram that meets all these requirements, as illustrated in Fig.~\ref{fig:4}. As sketched in the inset of Fig.~\ref{fig:4}a, zero field skyrmions are local energy minima bounded by two annihilation energy barriers $E_a^0$ and $E_a^\infty$ that prevent shrinking to zero size and infinite growth, respectively. By subtracting the internal deformation energy $2Ad$ from the minimum of $E_a^0$ and $E_a^\infty$, we obtain the effective annihilation energy barrier $E_a^\text{eff}$ that can be used to estimate long term thermal stability. For properly chosen material parameters, $E_a^\text{eff}$ can exceed the threshold $40k_BT$ (Fig.~\ref{fig:4}b) required for commercial storage devices. The corresponding radius is $<\SI{20}{nm}$. Note that all energy contributions generally become larger at larger radii. Therefore, fluctuations of any material parameter or of the field affect the energy barrier $E_a^\infty$ much stronger than $E_a^0$, see inset of Fig.~\ref{fig:4}a. This is why $E_a^\text{eff}$ increases slowly for increasing DMI values before it drops very rapidly after passing the maximum of $E_a^\text{eff}$, see inset of Fig.~\ref{fig:4}b. It is therefore advantageous to have a DMI value slightly below the maximum $E_a^\text{eff}$ to ensure that $E_a^\text{eff}$ is robust against moderate external fields. Finally, all zero field skyrmions have $E>0$, consistent with earlier assessments excluding ground state zero field skyrmions below the Curie temperature \cite{rosler_spontaneous_2006,wright_crystalline_1989}. It is reasonably easy to obtain $E_a^\text{eff}=10Ad$, which explains why the largest energy barriers can be obtained when increasing $A$ (or $d$).

\section*{The full phase diagram}

\begin{figure}
\centering
\includegraphics[width=\textwidth]{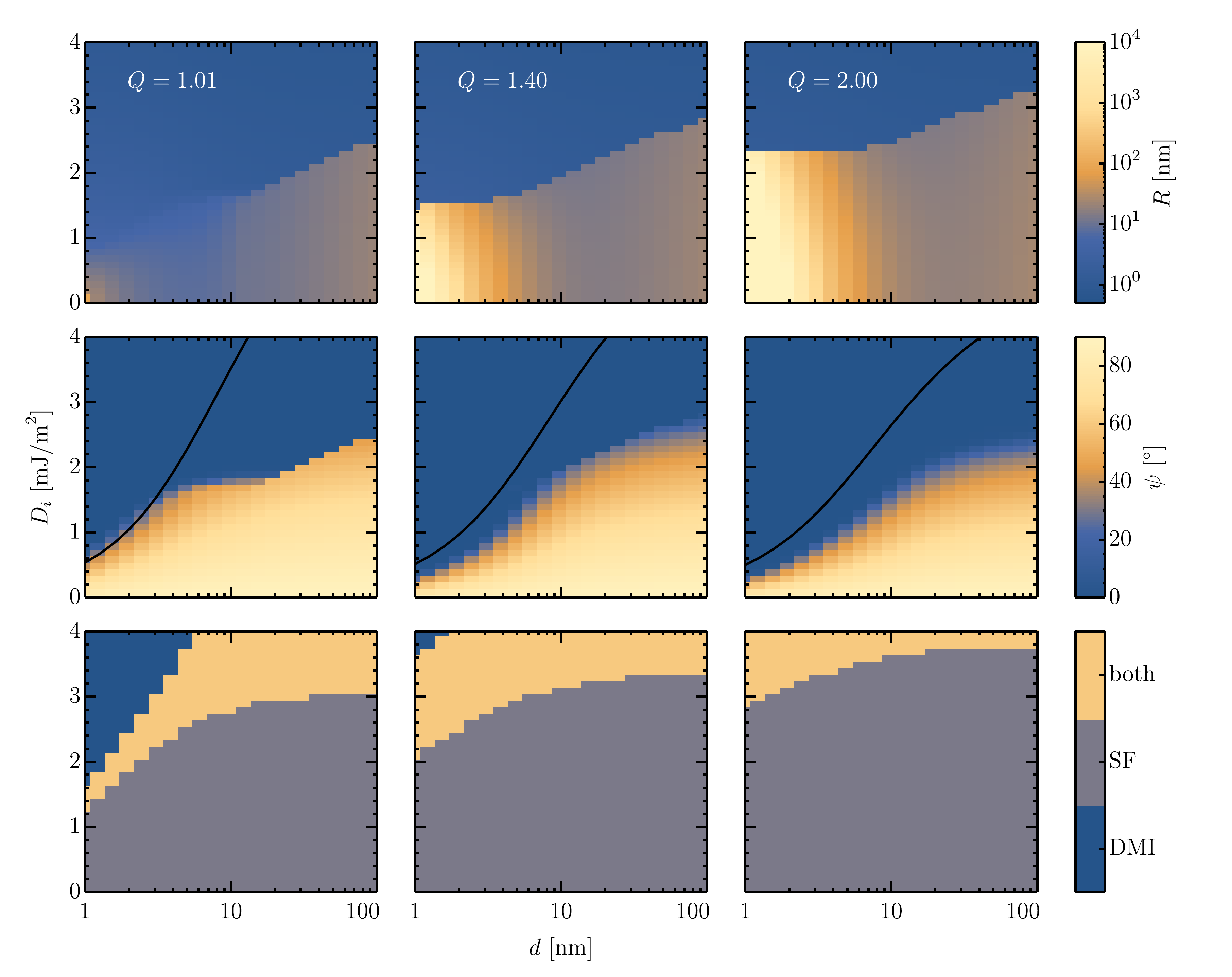}
\caption{\textbf{Skyrmion phase diagram.} Phase diagram as a function of total magnetic layer thickness $d$, interfacial DMI strength $D_i$, and anisotropy quality factor $Q$ for a multilayer with three layers of non-magnetic material for every layer of magnetic material. The panel rows show, from top to bottom: the radius $R$ and the DW angle $\psi$ of the smallest skyrmion just before collapse, and the possible means of stabilization (where ``DMI'' indicates that all skyrmions in this material are DMI stabilized, ``SF'' indicates that all skyrmions are stray field stabilized and ``both'' means that both types of stabilizations can be found, depending on the external field).  The solid line in the $\psi$ panels depicts the minimum DMI strength $D_{c\psi}^\text{SW}$ required to obtain fully N\'eel skyrmions ($\psi=0$) in all skyrmions, independent of their size.}
\label{fig:3}
\end{figure}

To demonstrate the power of our model and to understand the effect of material parameters on skyrmion properties, we derived and analyzed the properties of skyrmions as a function of more than one million different material parameters and magnetic fields, a task that is impossible with existing theoretical tools. Fig.~\ref{fig:3} illustrates some of the most interesting features of the derived full phase diagram. Specifically, the figure analyzes a magnetic multilayer in which the non-magnetic spacer layers (e.g., Pt and Ta) are three times as thick as the magnetic layers. All room-temperature skyrmion systems today are based on such multilayers \cite{moreau-luchaire_additive_2016,woo_observation_2016,litzius_skyrmion_2016,jiang_direct_2016,jiang_blowing_2015,buttner_dynamics_2015}. We employ the effective medium approach \cite{woo_observation_2016,lemesh_accurate_????} to treat these multilayers. The total magnetic material thickness in such films is between \SI{1}{nm} and \SI{100}{nm}, the interfacial DMI strength is below \SI{4}{mJ/m^2} and the anisotropy quality factor $Q=\frac{2K_u}{\mu_0M_s^2}$ is typically between 1 and 2. For this parameter range, we show in the first two rows of Fig.~\ref{fig:3} the radius $R$ and the DW angle $\psi$ of the smallest possible skyrmions, i.e., under the maximum field just before collapse. Below, we present the means of stabilization (DMI or stray fields). Similar diagrams with variable $A$, $M_s$, and non-magnetic spacer layer thickness are provided in the supplemental information.

The most striking common feature of all panels in Fig.~\ref{fig:3} is a clear phase boundary, i.e., a (thickness-dependent) critical DMI value above which the displayed quantity changes abruptly. For instance, for $D_i>D_{cr}$, skyrmions are extremely small ($\sim\SI{1}{nm}$) near collapse. At slightly lower DMI the skyrmion collapse radius can abruptly increase to several micometers. Note that this $D_{cr}$ has nothing to do with the DMI value for which the energy of an isolated straight wall becomes negative. The DW angle $\psi$ shows a qualitatively similar but quantitatively unrelated trend: Above a critical DMI value of $D_{c\psi}$, skyrmions are of N\'eel type when they collapse. The transition in $\psi$ is not as sharp as for $R$ and, importantly, $D_{c\psi}$ is consistently smaller than $D_{cr}$, implying that extremely small skyrmions are always of N\'eel type. Note also that small skyrmions are more likely to be of N\'eel type than straight walls in the same material. In other words, the critical DMI value for finding isolated N\'eel walls $D_{c\psi}^\text{SW}$ is much larger than $D_{c\psi}$. The region between $D_{c\psi}$ and $D_{c\psi}^\text{SW}$ is where the dependence of $\psi$ on the skyrmion size is most pronounced and bi-stable states are most likely to have different spin orientations.

Sub-\SI{10}{\micro m} skyrmions exist in almost the entire phase diagram. Materials with purely DMI stabilized skyrmions exist, but almost exclusively at very small values of $Q$. Already at $Q=1.4$, which is a typical value for cobalt based multilayers \cite{woo_observation_2016,litzius_skyrmion_2016}, purely DMI stabilized skyrmions exist only for DMI values larger than \SI{4}{mJ/m^2}, well beyond experimentally-reported values.  Hence we conclude that most skyrmions investigated experimentally so are best described as stray field stabilized.

Finally, comparing the additional phase diagrams in the supplemental information, we can qualitatively note that small skyrmions are favored by a low anisotropy, a low $M_s$, a small exchange constant, a large DMI value, and sizable non-magnetic spacer layers. Low $M_s$, low $A$ and thick spacer layers also lead to more abundant bi-stability regions, but note that here larger $Q$ values are beneficial.

\section*{Conclusions and outlook}

In summary, we presented an analytical model that allows exploration of the entire static phase diagram of isolated magnetic skyrmions via rapid, systematic calculations. We expect many new applications to arise from the exotic states found here, beyond what we already suggested. In principle, our model assumes infinite films and the behaviour in finite sized elements can be different \cite{rohart_skyrmion_2013}. However, in most cases confinement increases the stability of skyrmions, as long as skyrmions still fit into the element. Therefore, our predictions can be considered conservative and applicable to most nanostructures as well. Also, skyrmions in anti-ferromagnets \cite{barker_antiferromagnetic_2015,zhang_antiferromagnetic_2016} are covered by our theory by setting $M_s$ to zero. Still, some open challenges remain. For instance, the dynamics of skyrmions, and the effects of in-plane fields, are not yet covered by our model. But we believe that the concepts presented here to solve the integral equations pave the path to tackle those issues as well. 

\section*{Acknowledgements}

This work was supported by the U.S. Department of Energy (DOE), Office of Science, Basic Energy Sciences (BES) under Award \#DE-SC0012371. FB thanks Alexander Stottmeister, Benjamin Kr\"uger, and Kai Litzius for fruitful discussions and the German Science Foundation for financial support under grant number BU 3297/1-1.

\end{document}